\documentclass[a4paper,12pt]{iopart}

\usepackage{graphicx}

\begin{document}

\letter{Generalised Wigner surmise for (2$\times$2) random  matrices}
\author{P~Chau Huu-Tai$^\dag$, N A Smirnova$^\ddag$ and P~Van Isacker$^\dagger$}
\address{
 $^\dag$ Grand Acc\'el\'erateur National d'Ions Lourds, B.P. 55027, F-14076 Caen Cedex 5,
 France. \\
 $^\ddag$ University of Leuven, Instituut voor Kern- en Stralingsfysica,
 Celestijnenlaan 200D, B-3001 Leuven, Belgium.}

 \eads{ \mailto{chau@ganil.fr}, \mailto{nadya.smirnova@fys.kuleuven.ac.be} and
 \mailto{isacker@ganil.fr}}


\submitto{\JPA}
\pacs{05.40.-a, 05.30.Jp, 21.10.-k}

\begin{abstract}
 We present new analytical results concerning the spectral distributions for
 ($2\times 2$) random real symmetric matrices which generalise the Wigner surmise.
\end{abstract}

\section{Introduction}

 The level statistics of a quantum system represents the most significant,
 although not the only~\cite{Zel96}, signature of quantum chaos.
 The Poisson and Wigner distributions of dimensionless nearest-neighbour spacing, $s$,
 \begin{equation}
 \label{nnsp}
 \tilde{p}_P(s)= \exp{(-s)} \; ,
 \end{equation}
 \begin{equation}
 \label{nnsw}
 \tilde{p}_W(s)=\frac{\pi s}{2}
 \exp{\left(-\pi s^2/4\right)} \; ,
 \end{equation}
 are known in quantum chaos theory as two universalities
 that correspond to two extreme cases of classical dynamics, namely purely regular
 and  completely chaotic (see , e.g., \cite{Bohigas}).
 The majority of many-body systems such as
 nuclei, molecules, atoms or solids (see ~\cite{Zel96,Brod81,Guhr98} and references therein)
 have been found to be chaotic
 although for such complex systems  no classical limit can be constructed.

 As was first recognised by Wigner~\cite{Wigner58}, the  nearest-neighbour spacing distribution
 (NNSD) (\ref{nnsw})
 well corresponds to the eigenvalue distributions of random matrices,
 and this explains the importance of the Random Matrix Theory~\cite{Mehta90}
 for studying statistical properties of many-body systems.
 Particular attention  was paid to Gaussian ensembles.
 In fact, assuming that (i) the elements of the Hamiltonian matrix
 are {\em independent real variables} and
 (ii) the matrix distribution is {\em invariant}
 under an {\em orthogonal} transformation of the basis states
 (see, e.g., chapter 3 in \cite{Mehta90}),  one finds that the matrix elements are independent
 Gaussian variables with zero mean and with variance satisfying the conditions
 $\sigma^2_{ij}= (1+ \delta_{ij}) \sigma^2$.
 Imposing particular symmetries on the Hamiltonian, one gets~\cite{Dyson}
 Gaussian Orthogonal, Unitary or
 Symplectic Ensembles (GOE, GUE or GSE, respectively), which are widely and successfully
 applied in many fields of physics (see e.g. reviews~\cite{Zel96,Brod81,Guhr98}).

 Up to now, excited atomic nuclei are considered to be the best examples of chaotic quantum
 systems~\cite{Zel96}.
 Starting from slow-neutron scattering experiments, which were first described in terms
 of random matrices~\cite{Wigner58}, a lot of nuclear structure data has been analysed
 in the context of chaos.
 However, it was repeatedly noticed that the experimental data do not
 exactly match the distribution (\ref{nnsw}) and exhibit slight deviations~\cite{Bro73,NPA76}.
 These deviations are thought to be caused by the fact that the real system is not purely chaotic,
 but can be the quantum analog of a classical system that is  transitional  between
 chaotic and integrable.
 In this context, a few phenomenological formulae
 were proposed and analysed.
 The most famous are the Brody~\cite{Bro73} distribution,
 \begin{equation}
 \label{nnsb}
 \tilde{p}_{\omega }(s)=(\omega+1) \alpha s^{\omega }
 \exp{\left(-\alpha s^{\omega +1}\right)} \; ,
 \quad \alpha=\left[\Gamma \left(\frac{\omega+2}{\omega+1}\right)\right]^{\omega+1} \; ,
 \end{equation}
 which was shown to match better the experimental data~\cite{NPA76} on both high- and
 low-energy nuclear spectra,
 and the Berry-Robnik distribution~\cite{BeRo84}
 \begin{equation}
 \label{nnsbr}
 \tilde{p}_{BR}(s)=e^{(q-1)s}\{ (1-q)^2 \mbox{erfc}(\sqrt{\pi}
 q s/2)+[2q(1-q) + (\pi/2) q^3s] e^{-(\pi/4)q^2 s ^2}\}\; .
 \end{equation}
 Although the Brody distribution takes the form of Poisson
 for $\omega =0$ and Wigner for $\omega =1$, it has the unrealistic property that its derivative
 goes to infinity at $\varepsilon =0$ \cite{CaGr90}. Moreover, the parameter $\omega$ has no
 clear physical meaning.
 On the other hand, the distribution (\ref{nnsbr}) does not give a level-repulsion for non-integrable systems.
 Caurier et al~\cite{CaGr90} considered a model
 which allowed to simulate the transition
 from integrability to chaos and succeeded to derive the asymptotic limits
 for small and large neighbour spacings in the near-integrable limit.

 The idea to derive a third universality class corresponding to intermediate statistics
 has been actively pursued in recent years.
 In particular, in Refs.~\cite{JaKh99,AuJa01},
 the distribution of $n$ nearest neighbours,
 \begin{equation}
 \label{nnssp}
 \tilde{p}^{(\beta )}(n,s)= \frac{(\beta +1)^{n(\beta +1)}}{\Gamma[n(\beta +1)]}
 s^{[n(\beta +1)-1]} \exp{\left[-(\beta +1) s \right]} \; ,
 \end{equation}
 was shown to be relevant to a certain class of exactly solvable models
 with nearest and next-to-nearest neighbour interactions, generalizing the
 results obtained earlier~\cite{GrJa98,BoGe99} on pseudointegrable billiards
 and the short-range Dyson models.
 Remarkably, the NNSD given by (\ref{nnssp}) with $n=1$ exhibits a level repulsion
 $\sim s^{\beta }$ and falls to zero at $s$ as $\exp{[-(\beta +1)s]}$.
 For $\beta=1$ it is referred to as the semi-Poisson
 distribution \cite{GrJa98,BoGe99}.

 In this letter we argue that one might search for an explanation of the discrepancy between data and
  random matrix theory simply by generalising the random matrix ensemble used.
 In fact, the Gaussian ensemble is defined by the two assumptions (i) and (ii) mentioned above,
 and their applicability should be carefully checked for a given physical system.
 As far as nuclear physics is concerned,
 the deviations of the experimental data on slow-neutron or $(p,p')$
 resonances from a random matrix description
 could simply arise from the {\em non-invariance}
 of the random matrix ensemble under an orthogonal transformation of a basis, i.e.
 the assumption (ii) is violated.

 In addition, it is well known that realistic interactions in many-body nuclear,
 molecular or atomic systems are predominantly of  one- and two-body nature,
 implying that the distribution of the matrix is not only {\em not invariant}
 under an orthogonal (unitary) transformation of the basis, but also
 that the elements of the Hamiltonian matrix are {\em not independent}
 (if the number of particles is more than two), i.e. both assumptions
 for a Gaussian ensemble do not hold any longer.
 In this context, French and Wong~\cite{Fren70} and Bohigas and Flores~\cite{Bohi71}
 independently introduced the Two-Body Random Ensemble (TBRE), which is characterised by a
 Gaussian level density distribution,
 rather than a semi-circle provided by a GOE~\cite{Fren70}.
 The level distributions of experimental nuclear spectra did confirm this result.
 The NNSD relevant for a TBRE was found numerically to be fairly
 represented by the Wigner surmise~\cite{FrWo71},
 although recently it has  been pointed out~\cite{Jain01} that NNSD given by
 (\ref{nnssp}) with $n=1$
 and a certain real value of $\beta $ fits better the shell--model spectrum
 obtained with the $sd$-interaction of Wildenthal~\cite{USD}.

  Given the importance and actuality of these investigations,
  we present in this letter some analytical results concerning the properties of
 (2$\times$2) random symmetric matrices for which the assumption (ii)
 mentioned above is not satisfied.
 First, we derive the Hamiltonian distribution as a function of its eigenvalues
 and we calculate the NNSD which
 {\em generalises} the well-known Wigner surmise~\cite{Wigner58}.
 We show that the model allows to describe the transition from purely chaotic
 to asymptotically nearly integrable limits, being different from the intermediate
 statistics mentioned above.
 Then, for a particular case, we give the analytical expressions for the lowest moments
 of this distribution.
 Finally, we propose a method to derive
 the moments of the eigenvalue distribution  without knowledge of an explicit expression
 for the distribution.

 \section{Generalised Wigner surmise}

 Let us consider a (2$\times$2) real symmetric matrix
\begin{equation}\label{hamiltonian}
 \begin{array}{cc}\displaystyle
 H=\left(
 \begin{tabular}{cc}
 $H_{11}$    & $H_{12}$\\
 $H_{21}$    & $H_{22}$\\
 \end{tabular}
 \right)\,,
 \end{array}
 \end{equation}
 whose elements are
 independent Gaussian variables with zero mean and variance $\sigma_{ij}^{2}$,
 and $H_{12}=H_{21}$. The probability density of the matrix $H$ is then given by
\begin{equation}\label{probadensity}
 p(H)=\frac{1}{(2\pi)^{3/2}\sqrt{\sigma_{11}^2\sigma_{12}^2\sigma_{22}^2}}
 \exp\left[ -\left(\frac{H_{11}^{2}}{2\sigma_{11}^2}+
 \frac{H_{12}^{2}}{2\sigma_{12}^2}+\frac{H_{22}^{2}}{2\sigma_{22}^2}\right)\right].
\end{equation}
Each  matrix $H$ can be diagonalised in an orthogonal basis and therefore $H= O^{t}DO$, with
 $$
 O=
\begin{array}{cc}\displaystyle
\left(
\begin{tabular}{cc}
$\cos \theta$    & $-\sin\theta$\\
$\sin \theta$    & $\cos\theta$\\
\end{tabular}
\right),
\end{array}
$$
and
$$
D=
\begin{array}{cc}\displaystyle
\left(
\begin{tabular}{cc}
$E_{\alpha}$    & $0$\\
$0$    & $E_{\beta}$\\
\end{tabular}
\right).
\end{array}
$$
 Similar to the case of GOE \cite{BM}, we find that in the general case $$
\begin{array}{cc}\displaystyle
\begin{tabular}{ccc}
 $H_{11}$&=& $E_{\alpha}\cos^{2}\theta+E_{\beta}\sin^{2}\theta$\\
 $H_{12}$&=& $\left(E_{\alpha}-E_{\beta}\right)\cos\theta\sin\theta$\\
 $H_{22}$&=& $E_{\alpha}\sin^{2}\theta+E_{\beta}\cos^{2}\theta$\\
\end{tabular}
\end{array}.
$$
 We deduce that the probability density expressed in terms of the eigenvalues and
 the angle $\theta$ is
 \begin{eqnarray}
 \label{distrigene}
 \fl p\left(E_{\alpha},E_{\beta}, \theta \right)&=&
 \frac{E_{\alpha}-E_{\beta}}{(2\pi )^{3/2}\sqrt{\sigma_{11}^2\sigma_{22}^2\sigma_{12}^2}}
 \exp\left\{-\frac{\left[E_{\alpha}\Sigma^2- (E_{\alpha}-E_{\beta})
 \left(\sigma_{11}^2 \cos^2\theta+
 \sigma_{22}^2 \sin^2\theta \right)\right]^2}
 {2\sigma^2_{11}\sigma^2_{22}\Sigma^2} \right\} \nonumber \\
 & &
 \displaystyle \exp\left[-\frac12(E_{\alpha}-E_{\beta})^2
 \left( \frac{\cos^2 (2\theta )}{\Sigma ^2} +
 \frac{\sin^2 (2\theta )}{4\sigma_{12}^2} \right)\right]
 \end{eqnarray}
 where $\Sigma^2=\sigma_{11}^2 +\sigma_{22}^2$ and
 $E_{\alpha}-E_{\beta}\ge 0$.

 The nearest-neighbour spacing distribution
 for the variable $\varepsilon=E_{\alpha}-E_{\beta}= D s$, with the mean spacing $D=\int \varepsilon
 \tilde{p}\left(\varepsilon\right) d\varepsilon$, is given by the following integral
 \begin{equation}
 \label{p}
 \tilde{p}\left(\varepsilon\right)=
 \int_{-\pi/2}^{\pi/2}\!d\theta \int_{-\infty}^{\infty}\!dE_{\alpha}
 \int_{-\infty}^{E_{\alpha}} dE_{\beta} \; p\left(E_{\alpha},E_{\beta}, \theta \right)
 \delta\left(\varepsilon-E_{\alpha}+E_{\beta}\right),
 \end{equation}
 from which we obtain
 \begin{equation}
 \label{nnsgene}
 \tilde{p}\left(\varepsilon\right) =
 \frac{ \varepsilon}{2\sqrt{\Sigma ^2 \sigma_{12}^2}}
 \exp\left[-\frac{\varepsilon^2\left(\Sigma ^2 +
 4\sigma_{12}^2\right)}{16 \Sigma ^2 \sigma_{12}^2}\right]
 I_0\left( \frac{\varepsilon^2\left(\Sigma ^2 - 4\sigma_{12}^2\right)}{16
 \Sigma ^2 \sigma_{12}^2}\right)
 \end{equation}
 where  $I_0$ is a modified Bessel function of the first kind.

 The expression (\ref{nnsgene}) looks like a Rayleigh-Rice distribution,
 well known in signal theory~\cite{Picin93}, except for the argument of $I_0$,
 which is not linear as in the usual Rayleigh-Rice distribution but quadratic.
 This is why we will refer to $\tilde{p}\left(\varepsilon\right)$
 as to a {\em quadratic} Rayleigh-Rice distribution.

 Let us consider a particular case when the diagonal matrix elements
 have the same variance $\sigma_{11}^2=\sigma_{22}^2$, which is $\chi $ times larger than
 the variance, $\sigma_{12}^2=\sigma^2$, of the non-diagonal matrix elements, i.e.
 $\chi=\sigma_{11}^2/\sigma_{12}^2$.
 Then the eigenvalue distribution (\ref{distrigene}) reduces to
 \begin{equation}\label{districhi}
 p_{\chi}(E_{\alpha},E_{\beta},\theta)=
 \frac{E_{\alpha}-E_{\beta}}{(2\pi \sigma^2)^{3/2}\chi }
 \exp\left[-\frac{E_{\alpha}^2+E_{\beta}^2+
 \frac14 (E_{\alpha}{-}E_{\beta})^2 (\chi {-}2)\sin^2 (2\theta )}{2\chi \sigma^2}\right],
 \end{equation}
 while for the nearest-neighbour spacing we get
 \begin{equation}
 \label{nnschi}
 \tilde{p}_{\chi}(\varepsilon)=\frac{\varepsilon}{\sqrt{2\chi}2
 \sigma^2}\exp\left(-\frac{\left(\chi+2\right)\varepsilon^2}{16\chi\sigma^2}\right)
 I_0\left( \frac{(\chi-2)\varepsilon^2}{16\chi\sigma^2}\right).
 \end{equation}
 For $\chi=2$, expression (\ref{nnschi}) reduces to the Wigner surmise.
 The distributions $\tilde{p}_{\chi}(\varepsilon)$
 are plotted in figure 1 for $\chi=1$, $\chi=2$ and $\chi=5$.

\begin{figure}[htbp]
\begin{center}
\includegraphics[height=8cm,width=\linewidth]{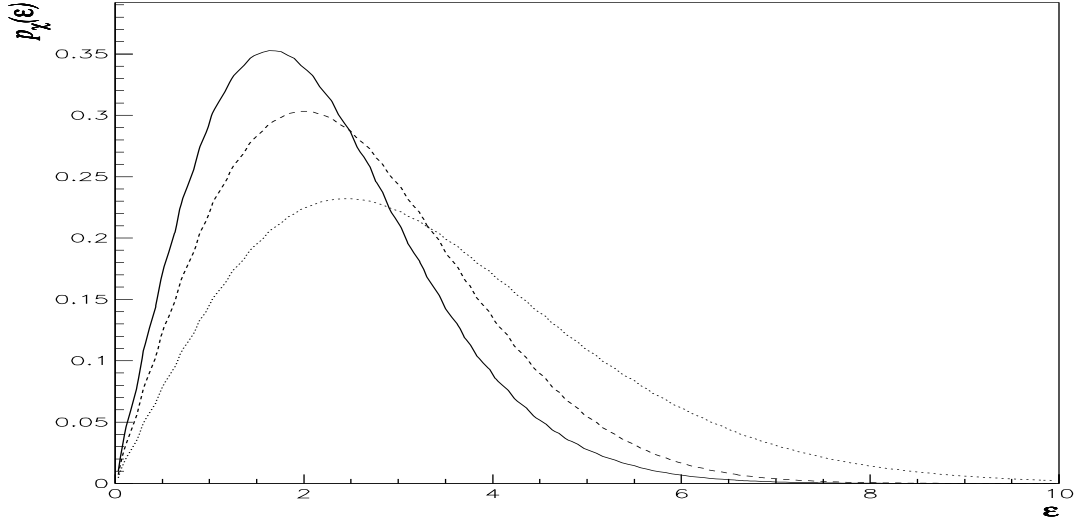}
\end{center}
\label{figRR}
\caption{Quadratic Rayleigh-Rice distributions for $\chi=1$ (solid line), $\chi=2$ (dashed line) and $\chi=5$ (dotted line)
with $\sigma^2=1$.}
\end{figure}

 For small values of $\varepsilon $, the distribution (\ref{nnschi})
 goes linearly to zero,
\begin{equation}
 \label{nns1}
 \tilde{p}_{\chi}(\varepsilon)\propto \frac{1}{\sqrt{2 \chi}2 \sigma^2} \varepsilon \; ,
 \end{equation}
 while for large $\varepsilon $,
 \begin{equation}
 \label{nns2}
 \tilde{p}_{\chi}(\varepsilon)\propto
 \left\{
 \begin{array}{ll}
 \displaystyle  \sqrt{\frac{2}{\mid\chi -2 \mid \sigma^2}}
 \exp\left(-\frac{\varepsilon^2}{4\chi\sigma^2}\right) & \chi \ne 2   \\
 \displaystyle \frac{\varepsilon }{4 \sigma ^2}
 \exp\left(-\frac{\varepsilon^2}{8\sigma^2}\right) & \chi=2 \\
 \end{array}
 \right.
 \end{equation}
 The asymptotic behaviour has a functional dependence on
 $\varepsilon $ similar to that of the Wigner surmise, i.e. the NNSD  (\ref{districhi})
 goes linearly to zero for $\varepsilon \to 0$ and it falls down according
 to $\exp{(-\varepsilon ^2)}$ for $\varepsilon \to \infty $.
 However, as seen from (\ref{nns1})--(\ref{nns2}),
 the natural dependence on $\chi $ provides a certain scaling.

 To calculate various statistical characteristics,
 it is often required to know certain moments of the distribution.
 Thus, we have derived analytical expressions for some moments of
 the distribution (\ref{nnschi}),
 and the lowest are given in \tref{tabMoment}.
\begin{table}[htbp]
 \caption{The moments up to $n=5$ of the quadratic Rayleigh-Rice distribution
 (\protect\ref{nnschi}) for $\sigma^2=1$. For $n=1$, one obtains the mean spacing $D$.
 $F$ is the hypergeometric function \cite{Huse97}.}
\lineup
\begin{indented}
\item[]\begin{tabular}{@{}lcll}
\br
& $M_n= \int_0^{\infty}\varepsilon^{n}\tilde{p}_{\chi}(\varepsilon)d\varepsilon$& values for $\chi=2$& for $\chi=5$\\
\mr
 $ n=0$ & 1& \0\01 &\0\0\01\\
 $ n=1$ &
 $4\chi\sqrt{2\pi}(\chi+2)^{-3/2}F\left[\frac{3}{4},\frac{5}{4},1,\left(\frac{\chi-2}{\chi+2}\right)^2\right]$&
 $\0\0\0\sqrt{2\pi} $ &\0\0\01.3 $\sqrt{2\pi}$ \\
 $ n=2$ & $\02\left(2+\chi\right)$& \0\08 &\0\014\\
 $ n=3$ &
 $96\chi^2\sqrt{2\pi}(\chi+2)^{-5/2}F\left[\frac{5}{4},\frac{7}{4},1,\left(\frac{\chi-2}{\chi+2}\right)^2\right]$
 & $\012\sqrt{2\pi} $
 &\0\028.7$\sqrt{2\pi}$ \\
 $ n=4$ & $4\left(12+4\chi+3\chi^2\right)$ & 128& \0428\\
$ n=5$ &
$3840\chi^3\sqrt{2\pi}(\chi+2)^{-7/2}F\left[\frac{7}{4},\frac{9}{4},1,\left(\frac{\chi-2}{\chi+2}\right)^2\right]$&$
240\sqrt{2\pi} $
&$ 1139.2\sqrt{2\pi}$ \\
\br
\end{tabular}
\end{indented}
\label{tabMoment}
\end{table}


From (\ref{nnschi}) and the expression of the mean spacing, $D$ (cf \tref{tabMoment}), we have derived the distribution of
the dimensionless nearest-neighbour spacing, $s=\varepsilon/D$ :
\begin{equation}
 \label{nnschired}
 q_{\chi}(s)= D \,\tilde{p}_{\chi}(s D)\,.
 \end{equation}
 For $\chi=2$, one finds the Wigner distribution (\ref{nnsw}).
 It can be shown that for $\chi$ and $\chi^{\prime}=4/\chi$,
 the two functions $q_{\chi}$ and $q_{\chi^{\prime}}$ are equal.

 The distributions $q_{\chi}$ for $\chi=4$, $\chi=30$, $\chi=500$ and $\chi=1000$
 are plotted in figure 2(a-d) and compared to
 the Poisson (dashed), Wigner (dotted) and
 semi-Poisson (dashed-dotted) distributions.
 Note that the semi-Poisson distribution can be fairly approximated for $\chi=30$
 although its asymptotic behaviour for large $s$ is different.
 As follows from figure 2, the parameter $\chi $ allows to describe
 a transition of a quantum system from a completely chaotic limit ($\chi=2$) to
 a nearly integrable one ($\chi \to \infty $ or $\chi \to 0$). However, the
 integrability characterised by the Poisson distribution (\ref{nnsp}) can never be reached
 (compare with the model of Caurier et al~\cite{CaGr90}).

\begin{figure}[htb]
\begin{center}
\includegraphics[height=10cm,width=\linewidth]{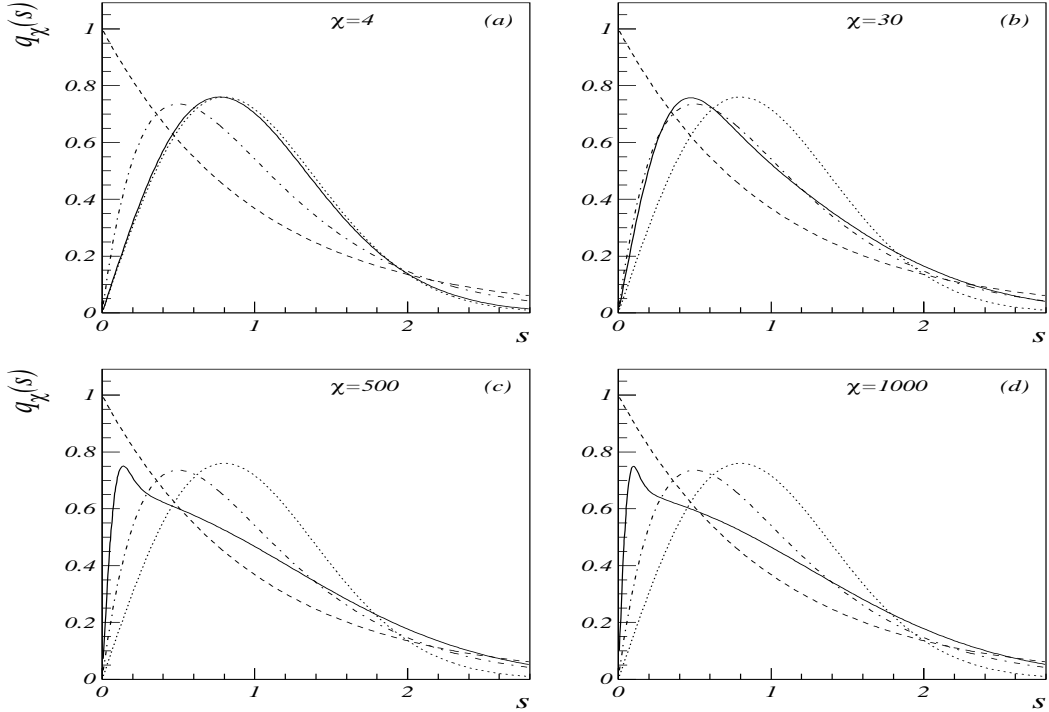}
\label{figDN}
\caption{Nearest neighbour spacing distributions  $q_{\chi}$ (solid line) for $\chi=4$ (a), $\chi=30$ (b),
$\chi=500$ (c) and $\chi=1000$ (d). The Poisson distribution is plotted in dashed line, the Wigner
distribution in dotted line and the semi-Poisson in dashed-dotted line.}
\end{center}
\end{figure}

 Indeed, integrating $p_\chi$  in (\ref{districhi}) over $E_{\beta }$ from $-\infty$ to $E_{\alpha }$
 and over $E_{\alpha}$ from $-\infty$ to $+\infty$, we obtain the
 angular distribution
 \begin{equation}\label{distritheta}
 r_{\chi}\left( \theta\right) = \frac{1}{\pi}\sqrt{\frac{\chi}{2}}\frac{1}{\left[1+\frac{1}{2}
 \left(\chi-2  \right)\sin^2\left(2\theta\right) \right]}.
 \end{equation}
 This distribution is represented in figure 3 for different values of $\chi$.
 For $\chi=2$ it is an exactly uniform distribution,
 which means that there is no privileged basis (the orthogonal invariance holds).
 For high values of $\chi$, the initial basis is nearly
 the eigenbasis (the diagonal elements are much larger than the non-diagonal ones),
 thus $r_{\chi}$ takes its maximum absolute values for
 $\theta = 0$ and $\theta = \pi /2 $, whereas for small values of $\chi$,
 the eigenstates are more likely obtained after
 a rotation of $\pi/4$ of the initial basis and $r_{\chi}$ is maximum for
 $\theta =\pi /4 $. For $\chi$ and $\chi^{\prime}=4/\chi$ the two curves are in quadrature.

\begin{figure}[htbp]
\begin{center}
\includegraphics[height=8cm,width=11cm]{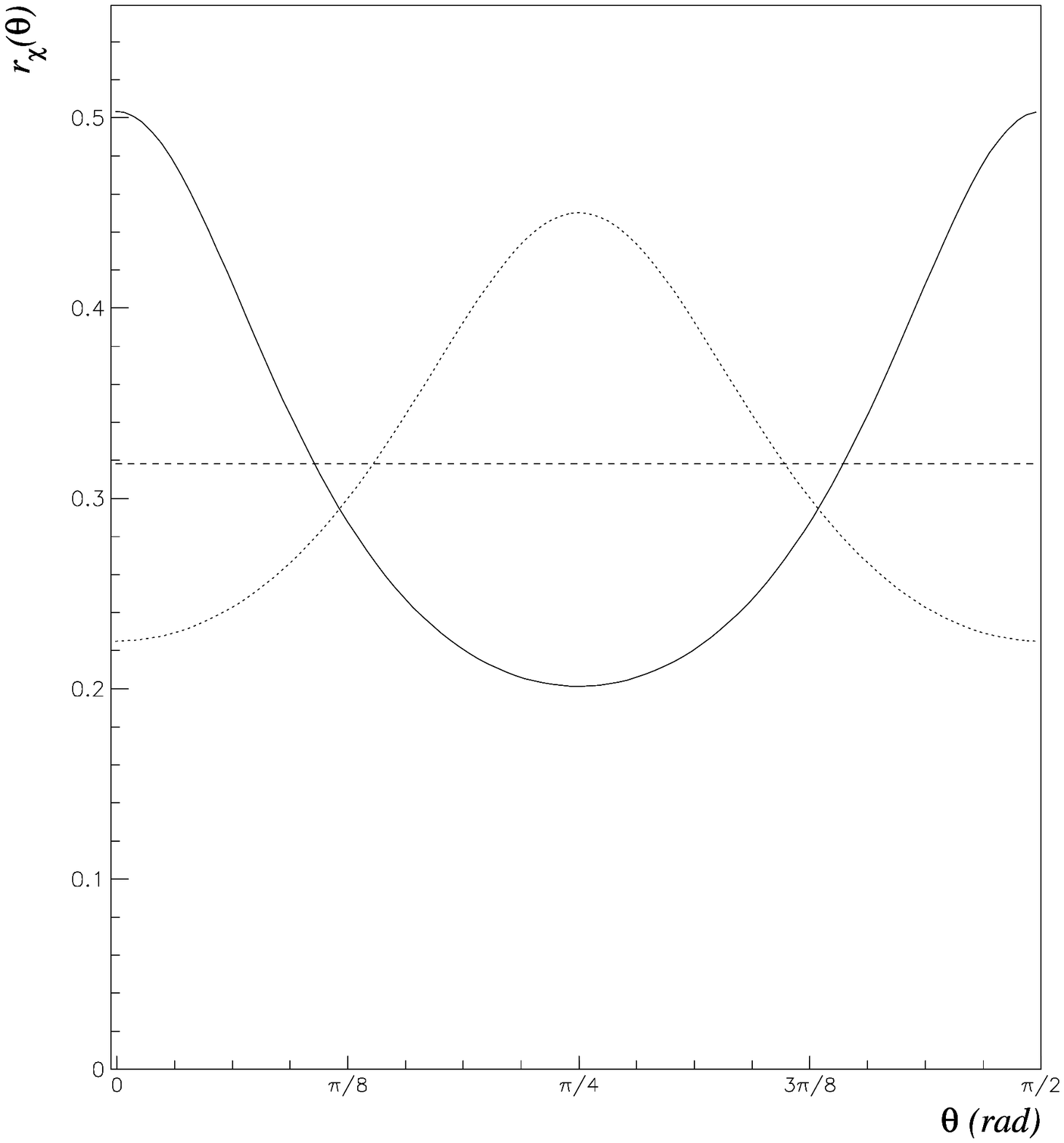}
\end{center}
\label{figTheta}
\caption{Angular distributions for $\chi=1$ (dotted line), $\chi=2$ (dashed line) and
$\chi=5$ (solid line).}
\end{figure}

 We can re-express $p_{\chi}(E_{\alpha },E_{\beta })$ as a function of $\varepsilon$ and
 $S=E_\alpha+E_\beta$. Then $p_{\chi}$ can be factorised
 into a function depending on $\varepsilon$ times a function depending on $S$,
 i.e. these variables are independent.
 Moreover since $S$ is the trace of the matrix it is a Gaussian variable with zero mean and
 all its odd moments are zero.
 From the independence of $\varepsilon$ and $S$ we deduce that the moments of the eigenvalues
 fulfill
 \begin{eqnarray}
 \langle E_{\alpha}^n\rangle & = &(-1)^n \langle E_{\beta}^n\rangle \label{moment1}\\
 \langle E_{\alpha}^n\rangle & = & \frac{1}{2^n}\sum_{p=0}^{n/2} \left(
 ^n_{2p}\right)
 \langle S^{2p}\rangle \langle \varepsilon^{n-2p}\rangle \label{moment2}
 \end{eqnarray}
 From  (\ref{moment1}) and (\ref{moment2}),
 we can derive the moments of the eigenvalues whose distributions are
 difficult to compute.
 Note that  we deduce from (\ref{moment1}) that the highest  and
 the lowest eigenvalues have opposite mean values and the same variance.

\section{Conclusion}

 The study of the statistical properties of spectra of realistic
 Hamiltonians requires the consideration of a random matrix ensemble whose elements are
 not independent or whose distribution is not invariant under orthogonal
 transformation of a chosen basis.
 In this letter we have concentrated on the properties of
 (2$\times$2) real symmetric matrices whose elements are independent Gaussian variables
 with zero means but do not belong to the GOE.
 We have derived the distribution of eigenvalues for such a matrix,
 the NNSD which generalises the Wigner surmise and
 we have calculated some important moments.
 The asymptotic properties of the distribution obtained are functionally identical
 to those of the ordinary Wigner surmise.
 For finite $\chi $, the model considered here allows to describe
 the transition from chaos to near integrability
 (the exact integrable limit is never realized).
 Thus it represents a chaotic system although with a degree of disorder
 less important than in the Wigner surmise.
 The derivation of similar analytical expressions for matrices of larger
 dimensions is technically difficult. However, we believe that the present results already justify
 the use of NNND of type (\ref{nnschi}) to fit the data as an alternative to the Brody distribution.
 We also think that these results can be extended to hermitian matrices.

\section*{Acknowledgements}

 We would like to thank Olivier Juillet for his help to calculate some integrals and
 Alejandro Frank for his helpful and enthusiastic discussions.
 PCH-T would like to thank Sylvie Hudan for her help to draw the figures.
 NAS is a post-doctoral researcher of the FWO-Vlaanderen, Belgium.

\section*{References}

\end{document}